\documentclass[twocolumns]{aastex63}

\graphicspath{{./}{figures/}}
\begin{document} 

\title{The importance of horizontal Poynting flux in the solar photosphere}

\correspondingauthor{Suzana S. A. Silva}
\email{suzana.silva@sheffield.ac.uk}

\author[0000-0002-0786-7307]{Suzana S. A. Silva}
\affiliation{Plasma Dynamics Group, Department of Automatic Control and Systems Engineering, University of Sheffield, Sheffield, UK
 \\}
 \affiliation{Department of Physics, Aeronautics Institute of Technology, S\~{a}o Jos\'{e} dos Campos, Brazil \\
}

\author{Mariarita Murabito}
\affiliation{INAF – Osservatorio Astronomico di Roma, Via Frascati,33 Monte Porzio Catone, RM, 00078, Italy\\}

\author[0000-0002-7711-5397]{Shahin Jafarzadeh}
\affiliation{Max Planck Institute for Solar System Research, Justus-von-Liebig-Weg 3, 37077 G\"{o}ttingen, Germany\\}
\affiliation{Rosseland Centre for Solar Physics, University of Oslo, P.O. Box 1029 Blindern, NO-0315 Oslo, Norway\\}

\author{Marco Stangliani}
\affiliation{ASI, Italian Space Agency, Via del Politecnico snc, 00133, Rome, Italy\\}

\author{Gary Verth}
\affiliation{Plasma Dynamics Group, School of Mathematics and Statistics, University of Sheffield, Sheffield, UK
 \\}

\author{Istvan Ballai}
\affiliation{Plasma Dynamics Group, School of Mathematics and Statistics, University of Sheffield, Sheffield, UK
 \\}

\author{Viktor Fedun}
\affiliation{Plasma Dynamics Group, Department of Automatic Control and Systems Engineering, University of Sheffield, Sheffield, UK
 \\}

\begin{abstract}
 The electromagnetic energy flux in the lower atmosphere of the Sun is a key tool to describe the energy balance of the solar atmosphere. Current investigations on energy flux in the solar atmosphere focus primarily on the vertical electromagnetic flux through the photosphere, ignoring the Poynting flux in other directions and its possible contributions to local heating. Based on a realistic Bifrost simulation of a quiet-Sun (coronal hole) atmosphere, we find that the total electromagnetic energy flux in the photosphere occurs mainly parallel to the photosphere, concentrating in small regions along intergranular lanes. Thereby, it was possible to define a proxy for this energy flux based on only variables that can be promptly retrieved from observations, namely, horizontal velocities of the small-scale magnetic elements and their longitudinal magnetic flux. Our proxy accurately describes the actual Poynting flux distribution in the simulations, with the electromagnetic energy flux reaching  $10^{10}$ erg\,cm$^{-2}$\,s$^{-1}$.  To validate our findings, we extended the analysis to SUNRISE/IMaX data. First, we show that Bifrost realistically describes photospheric quiet-Sun regions, as the simulation presents similar distributions for line-of-sight magnetic flux and horizontal velocity field. Second, we found very similar horizontal Poynting flux proxy distributions for the simulated photosphere and observational data. Our results also indicate that the horizontal Poynting flux in the observations is considerably larger than the vertical electromagnetic flux from previous observational estimates. Therefore, our analysis confirms that the electromagnetic energy flux in the photosphere is mainly horizontal and is most intense in localized regions along intergranular lanes.
\end{abstract}


\section{Introduction}
Essential aspects of energy generation, transport, and dissipation responsible for the heating of the upper atmosphere of the Sun remain elusive. Although the ongoing turbulent motion in the upper convective zone, particularly in the intergranular lanes \citep{Pontin_2020}, can be considered a natural reservoir of energy for heating the upper solar atmosphere, the question remains of how this energy can be transferred to the higher regions.  
By shuffling the magnetic field lines, the convective motions would be able to increase sufficiently the magnetic energy to compensate for coronal and chromospheric energy losses, driving injection of magnetic energy later transformed into heating by AC or DC mechanisms. The magnetic nature of the solar atmosphere's dynamics and the magnetic coupling between different heights indicate the flux of electromagnetic energy, i.e. the Poynting flux, as the primary mechanism for energy transport \citep{Norlund_2009,Shelyag_2012}.

The electromagnetic energy flux input in the upper solar atmosphere is described by the vertical component of the Poynting flux produced by vertical plasma motions transporting horizontal magnetic field and the work done by horizontal motions on the vertical magnetic field \citep{Shelyag_2012}. The importance of vertical and horizontal plasma velocities to the upward Poynting flux depends on the magnetic configuration of the solar atmosphere.  Based on realistic MHD simulations of the lower atmosphere, \cite{Steiner_2008} found that a positive vertical component of the Poynting flux in the photosphere could be provided by horizontal magnetic fields expelled to the upper photosphere by convective overshooting. For a simulated solar plage,  \cite{Shelyag_2012} found that the horizontal motions of the solar photospheric vortices would be the primary sources of vertical Poynting flux, encompassing regions with intense electromagnetic energy flux going up and down, but resulting in a total positive net. Besides playing an essential role in generating an upwards pointing Poynting flux, the small-scale vortices could provides enough energy to heat the chromosphere \citep{Yadav_2020}. The swirling motions in the chromosphere also provides a mean net upwardly directed Poynting flux and can spatially correlate to intense vertical Poynting flux distributions \citep{Battaglia_2021}.\\
In active regions, a strong correlation between the averaged X-ray brightness and a proxy for Poynting flux was first suggested by \cite{Fisher_1998}. \cite{Tan_2007} confirmed that correlation and found the Poynting flux to be in the range of $10^{6.7}$ - $10^{7.6}$ ergs {cm$^{-2}$} s$^{-1}$, which provides enough energy to justify coronal temperatures. According to estimates based on observational data, the upward Poynting flux can also provide the necessary energy input in solar plage regions \citep{Yeates_2014,Welsch_2015}

The Poynting flux input from the photosphere is not constant and tends to strongly vary as a function of time \citep{Shelyag_2012, Hansteen_2015}. The highly variable nature of the vertical Poynting flux results from the Poynting flux magnitude being regulated by the angle between the magnetic field and the solar surface, which tends to fluctuate as the magnetic field lines are dragged and twisted by photospheric motions. 
The dependency on magnetic energy and the plasma velocity transporting the energy is also one of the causes for the high variability of the Poynting flux. Thereby, a complete description of the Poynting vector in the photosphere is an essential aspect for an accurate description of the electromagnetic energy transport in the solar atmosphere. However, those dependencies of the Poynting flux make observational estimates highly dependent on both the methodology to recover photospheric velocity and the assumptions used to describe the 3D electromagnetic field in the photosphere \citep{Fisher_2012,  Kazachenko_2014, Welsch_2015}. Another challenge is that the analysis on solar energy transport only considered a narrow class of plasma motions, neglecting the complexity of plasma flows and their capacity to reorganize the magnetic field.  \cite{Sakaue_2020} found that intermediate shock regions can reduce transmission of the Poynting flux by Alfvén waves; therefore, it is not well established how much of the energy generated by photospheric motions will reach the chromosphere and solar corona. 

Based on realistic Bifrost simulation of a quiet-Sun atmosphere, our work focuses on {an approximate description} of the total Poynting flux for the wealthy dynamics of photospheric plasma flow. We show that the total Poynting flux can be  evaluated from an approximation that relies only on physical plasma variables that can be established from observations: horizontal plasma motions, as derived from tracking magnetic elements, and line-of-sight magnetic flux obtained from spectropolarimetric inversions. We extend our analysis to observations using {\sc Sunrise}/IMaX data and validate the simulation findings. Although our approximation for Poynting flux estimations has been previously applied to describe upward electromagnetic energy flux in active regions \citep{Fisher_1998, Tan_2007}, we demonstrate that such proxy, in reality, describes the horizontal Poynting flux with considerable accuracy. Moreover, our analysis clearly shows that the vertical electromagnetic energy flux is only a small fraction of the total Poynting flux. Our findings also indicate the upper photosphere and lower chromosphere as the regions with higher contributions for upward electromagnetic energy flux, suggesting those regions as the most likely energy sources in the upper atmosphere. Our paper is organized as follows: Section 2 describes the Bifrost simulation and the methodology to retrieve the velocity field from {\sc Sunrise}/IMaX observations; Section 3 presents the description of the photospheric Poynting flux, our approximation for observation of the Poynting flux, and the validation of our findings based on {\sc Sunrise}/IMaX data. Finally, Section 4 presents a discussion of our results and conclusions on our findings.

\section{Methodology}

\subsection{Bifrost Simulations}

The numerical data used in our study were obtained using the Bifrost code \citep{Gudiksen_2011}, which simulates realistic magnetoconvection in the solar photosphere and upper convection zone based on radiative MHD equations. The simulation domain consists of $768 \times 768 \times 768$ grid cells, covering 24 Mm in the
$x-$ and $y$-directions and 16.8 Mm in the vertical $z$-direction, with 31~km horizontal resolution and 12-82~km variable vertical resolution. The simulated atmosphere reaches up the lower corona, extending vertically to 14.3 Mm above the surface. The model used in the numerical approximation realistically mimics a coronal hole (i.e. a quiet-Sun region with open fields in the upper atmosphere), consisting of only small-scale opposite-polarity magnetic fields in the photosphere with an average signed field strength of 5~G and a mean unsigned field strength of 40 G. Thereby, the Bifrost data set adequately represents regions with predominantly vertical magnetic fields. The visible solar surface (Rosseland optical depth $\tau=1$) is located at 2.5~Mm above the lower boundary. At the bottom boundary, a horizontal field of 200~G along the y-axis was fed into the inflows in a way that the magnetic field strength is slowly increasing with time and there are interactions between the existing field and the flux emergence. {The top and bottom boundaries of the simulation box are transparent, and the} lateral boundaries are periodic. The radiative transfer equations in the photosphere and low chromosphere take into account scattering \citep{Skartlien_2000}, and the chromospheric (and transition region) radiative losses are computed in  non-LTE conditions \citep{Carlsson_2012}; Hydrogen ionization in this simulation is treated in LTE. For further details we refer to \cite{DePontieu_2021} where the same simulation run (with identity number ch024031\_by200bz005) has also been analyzed.

\begin{figure}[htp!]
    \centering
    \includegraphics[trim = 0mm 0mm 0mm 0mm, clip, width=1\columnwidth]{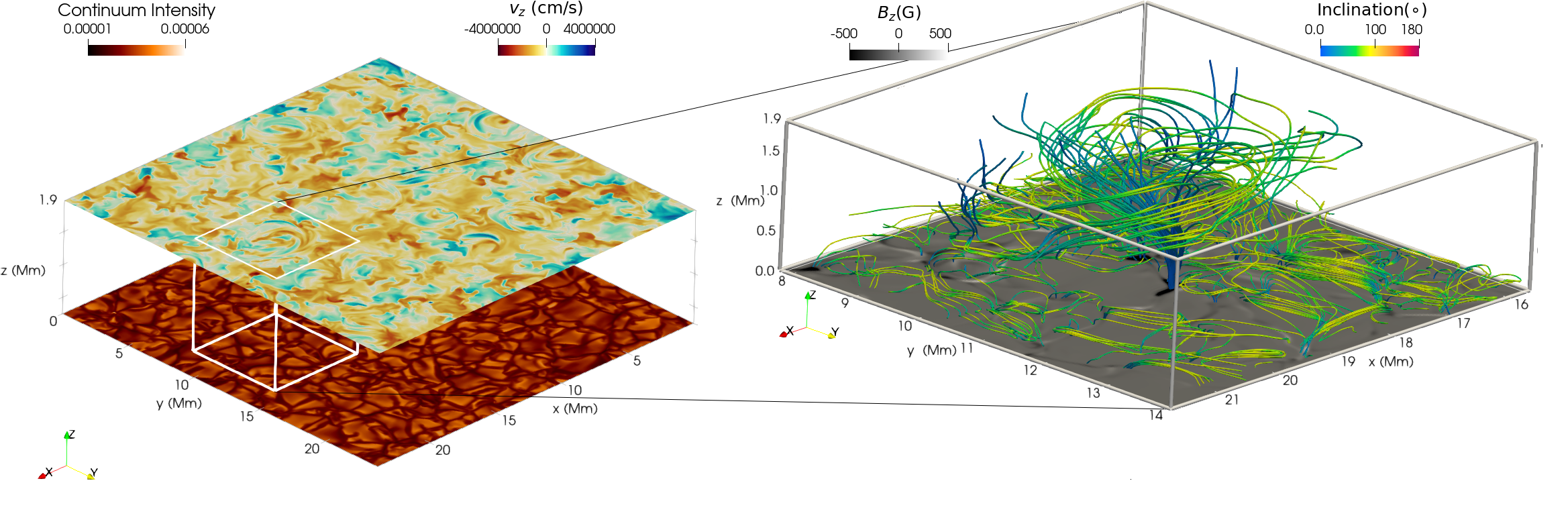}
    \caption{The Bifrost simulation domain. {Left panel}: the $(x,y)$ planes are placed 26 km above the simulated solar surface, colored by the continuum intensity, and 1.9 Mm above the surface, colored by the vertical component of the velocity field. The white cuboid shows the region used for close view of the simulated domain. {Right panel: depiction of the magnetic field lines for the region in the white cuboid from the left panel. The $(x,y)$ plane is placed 26 km above the simulated solar surface, colored by the $z$-component of the magnetic field saturated at 500 G. The magnetic field lines are colored by the inclination of the magnetic field.} }
    \label{fig:simulation_domain}
\end{figure}

Figure \ref{fig:simulation_domain} shows the simulated atmosphere at time 76' 46". In the left panel, we see the whole domain and the lower horizontal plane is placed 26 km above the simulated solar surface colored by the continuum intensity. The horizontal domain covers multiple granules and their intergranular regions, an important part of our Poynting flux analysis. In the upper horizontal plane, placed at 1.9 Mm above the surface, we see the distribution of the vertical velocity field, $v_z$. Although it is not the scope of our analysis, it is interesting to point out that these Bifrost data also simulate the chromospheric swirls, as indicated by the ring shapes formed in $v_z$. {The white cuboid shows the region used for close view of the simulated domain, and the magnetic field topology for that region is shown in the right panel of Fig. \ref{fig:simulation_domain}. The horizontal plane is placed 26 km above the simulated surface and is color-coded by the $z$-component of the magnetic field. The magnetic field lines are traced from random uniformly distributed points in the simulated photosphere, and they are colored by their inclination with respect to the vertical direction. The magnetic field lines display a wealthy topology, being mainly horizontal in the regions lying close to the surface and presenting twists and lower inclination at upper heights.}

\subsection{{\sc Sunrise}/IMaX observations}
To validate our analysis based on the simulation data set and estimate the Poynting flux in the photosphere, we extend our investigation to observational data. More specifically, a seeing-free spectropolarimetric dataset obtained by the Imaging Magnetograph eXperiment \citep[IMaX;][]{Martinez2011} on board the first flight of the {\sc Sunrise} balloon-borne solar observatory \citep{Solanki_2010,Barthol2011,Berkefeld_2011} is being analyzed. The data were acquired on 2009 June 9, and consist of 32 minutes (for a total of 100 scans) of full Stokes measurements along the Fe~{\sc i}~525.02~nm spectral line (in four wavelength positions in the line and one in the continuum) with a cadence of 33 s and a spatial sampling of 0.0545 arcsec/pixel. The data were phase-diversity (PD) reconstructed, reaching to a (single-position) noise level of $\approx3\times10^{-3}$ in the units of Stokes~$I$ continuum. The field of view (FOV) is approximately 40$\times$40 arcsec$^2$ and samples a quiet-Sun region at disk center.
In our study we used the results of the Stokes inversion {previously} presented by \citet{Kianfar2018}. In particular, the longitudinal component of the magnetic field strength, retrieved by the SPINOR inversion code \citep{Frutiger_2000,Berdyugina_2003}, was used to estimate the Poynting flux vector. For further details on the Stokes inversion procedure we refer the reader to \citet{Kianfar2018} and \citet{Kahil_2017}.

For our analysis, the photospheric horizontal velocity, $v_{h}$, was retrieved from the {\sc Sunrise}/IMaX observations based on tracking the magnetic elements detected in the data using the SWAMIS\footnote{available at https://www.boulder.swri.edu/swamis/} \citep{DeForest2007_swamis} code. 
The same analysis was also applied to the simulation data (at 115 km above the solar surface) in order to compare the results. 
The SWAMIS algorithm works by \begin{itemize}
    \item Detecting: magnetic elements are identified in each image.
    \item Identifying: the detected elements are given a unique identification number.
    \item Associating: the elements in one image are associated with the same elements in the preceding and following images.
\end{itemize}
In the case of {\sc Sunrise}/IMaX data, the code tracks and identifies groups of pixels in the circular polarization (CP) maps. For the simulation data, we convolved the vertical magnetic field ($B_{z}$) with a point-spread function (PSF) similar to the IMaX one, and we use this resulting map as input for the magnetic tracking.
The {\sc Sunrise}/IMaX circular polarization maps are constructed from the Stokes $V$ signals pixel-wise, following the method set out by \citet{Martinez2011}: 
\begin{eqnarray}
\mathrm{CP}=\frac{1}{4 \left \langle I_{c} \right \rangle} \sum_{i=1}^{4} \epsilon_ {i} {V_{i}}  
,\end{eqnarray} 
\noindent
where $\epsilon_{1,2}$ = 1 for the first two spectral positions of the line sampling (i.e. in the blue wing) and $\epsilon_{3,4}$ = -1 for the next two positions (i.e. in the red wing). In the above equation $I_{c}$ is the normalization factor applied to the Stokes $V$ profiles, and it is derived taking the Stokes $I$ continuum intensity (i.e. the fifth wavelength position) over each corresponding pixel. The CP has a considerably lower noise level of $1.7\times10^{-3}$~$I_{c}$ \citep{Jafarzadeh_2014}, and thus a larger signal-to-noise ratio (compared to the signal of the single Stokes $V$ wavelength positions), facilitating the detection of small-scale magnetic structures.

The SWAMIS code uses a two-threshold (high and low) signed discriminator. These thresholds are chosen as 2$\sigma$ and 6$\sigma$, respectively, where $\sigma$ corresponds to the standard deviation of the magnetic signal over the FOV for both the {\sc Sunrise}/IMaX's CP and $B_{z}$ simulation maps. 
Then, we selected suitable and representative magnetic elements on the basis of their lifetimes (longer than 10 consecutive frames) and sizes (minimum 8 pixels) and we estimated their instantaneous horizontal velocity, $v_{h}$, by tracking their positions in time.

\begin{figure}[htp]
    \centering
    \includegraphics[width=0.5\textwidth]{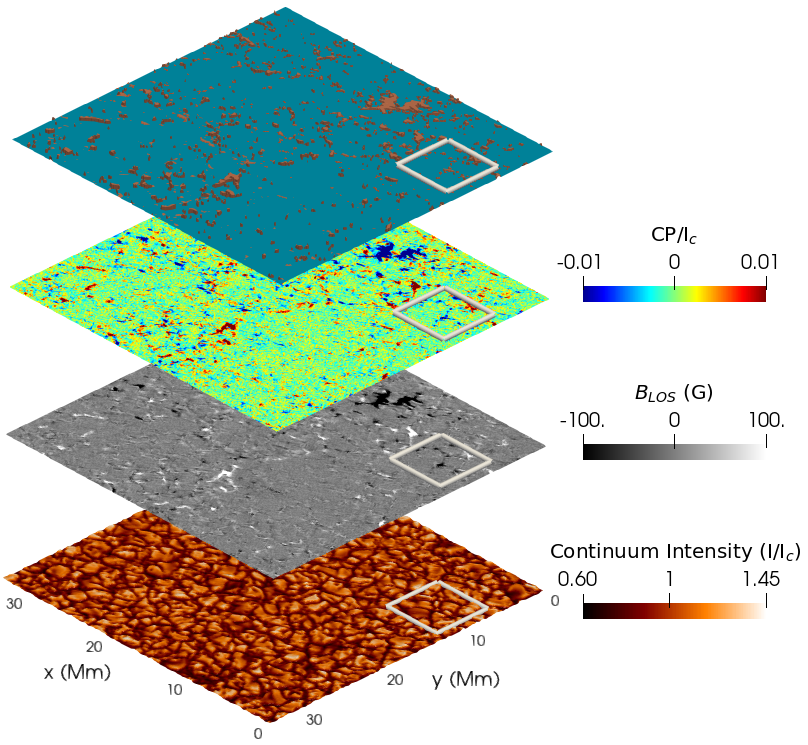}
    \caption{From bottom to top: IMaX continuum intensity, magnetic field obtained from SPINOR inversions saturated at 100 G, IMaX CP maps and  magnetic elements depicted in orange. The white box indicates the sub-FOV shown in Fig.~\ref{fig:deepvel}.}
    \label{fig:IMaX}
\end{figure}

The {\sc Sunrise}/IMaX data at photospheric heights are shown in Fig.~\ref{fig:IMaX}. In particular, this figure displays (from the bottom to the top) the continuum intensity, the magnetic field as retrieved by the SPINOR inversion, and the CP maps. The magnetic elements that fulfill the SWAMIS code's criteria are shown in orange at the top (x,y) plane. As shown by Fig.~\ref{fig:IMaX}, the code identifies all the pixels with a magnetic signal from the CP, which are direct measurements of the magnetic field along the line of sight.  

Finally, we employed the DeepVel method\footnote{available at https://github.com/aasensio/deepvel}  developed by \citet{Asensio2017} to estimate the instantaneous horizontal velocities, $v_{h}$, at all pixels over the entire FOV of the {\sc Sunrise}/IMaX data.  
Deepvel is a deep neural network that optimizes its network weights by considering the outputs and inputs from the two ends through convolution using a series of transformations that relates the transverse velocities' vector field (the output) to the maps of the intensity $I_c$ (input) at the surface. The Deepvel applied in this paper was already trained for the IMaX images with a 30~s cadence.

\section{Results}
\subsection{Electromagnetic energy flux in the solar atmosphere}
The electromagnetic energy flux is described by the magnetic Poynting vector, which can be written in MHD approximation as
\begin{equation}
    \mathbf{S} = \frac{1}{4 \pi}\mathbf{B}\times (\mathbf{v}\times \mathbf{B}),
    \label{eq:Pflux}
\end{equation}
where $\mathbf{v}$ is the plasma velocity and $\mathbf{B}$ is the magnetic field. To describe the {energy flux} orientation in relation to the magnetic field, we write the velocity field in terms of its parallel ($\mathbf{v}_{\parallel}$) and perpendicular ($\mathbf{v}_{\perp}$) components to $\mathbf{B}$. As a result, Eq.\ \ref{eq:Pflux} becomes
\begin{equation}
     \mathbf{S} = \frac{1}{4 \pi}  \mathbf{v}_{\perp}(\mathbf{B} \cdot \mathbf{B}).
     \label{eq:Pflux2}
\end{equation}
Therefore, the electromagnetic energy flux can be interpreted as magnetic energy density transported by the portion of the plasma velocity flowing perpendicular to the magnetic field. {The above expression includes the contributions from transverse waves, as the transverse perturbations in the magnetic field lead to creation of horizontal components of the magnetic field and thereby a vertical  component for the Poynting flux and thus energy transport in vertical direction. Equation \ref{eq:Pflux2} also indicates that the transport of electromagnetic energy to the upper atmosphere require magnetic field lines with  inclination in respect to the vertical direction. }

\begin{figure*}[ht!]
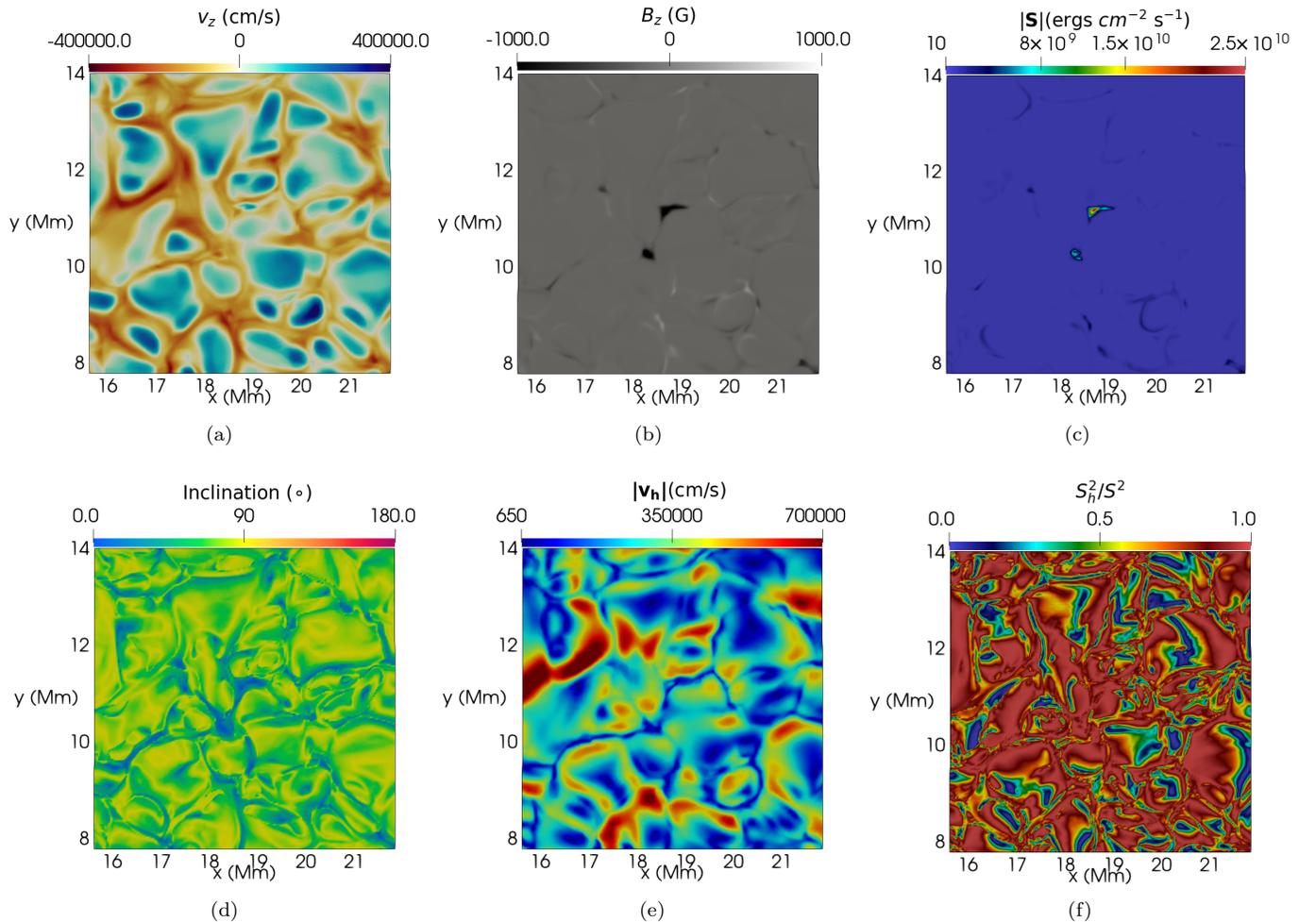

    \centering
\gridline{\fig{sim_vz.png}{0.33\textwidth}{(a)}
          \fig{sim_bz.png}{0.33\textwidth}{(b)}
\fig{sim_S.png}{0.33\textwidth}{(c)}}
\gridline{
          \fig{sim_inclination.png}{0.33\textwidth}{(d)}
          \fig{sim_vh.png}{0.33\textwidth}{(e)}  
                    \fig{sim_ratio.png}{0.33\textwidth}{(f)}  }
    \caption{Close view of a 6.25x6.25 Mm$^2$ region from the Bifrost simulation indicated by the white rectangle in Fig. \ref{fig:simulation_domain}. The horizontal planes are  placed $\sim 115$ Km above the solar surface. {The individual panels illustrate}: (a) the vertical component of the velocity field, (b) vertical component of the magnetic field, (c) the magnitude of the Poynting flux, (d) the inclination of the magnetic field, (e) the horizontal plasma velocity and (f) the ratio $S_h^2/S^2$. }
    \label{fig:uz+pf}
\end{figure*}

The first hint on electromagnetic energy flux in the lower atmosphere is provided in a close view, {$6.25\times6.25$} Mm, {of a horizontal plane at $\sim 115$ km above the simulated solar surface as displayed in Fig.~\ref{fig:uz+pf}.
The granular patterns of the region are indicated by the vertical component of the velocity field, $v_z$,  in Fig. ~\ref{fig:uz+pf}(a)}.
The vertical photospheric magnetic field is shown in Fig.~\ref{fig:uz+pf}(b), with  the color map saturated at $\pm$1000 G, indicating the magnetic flux concentration of both polarities along the intergranular lanes. Figure~\ref{fig:uz+pf}(c) displays the magnitude of electromagnetic energy flux generated by the photospheric dynamics. Comparing Figures ~\ref{fig:uz+pf}(a) and (c), we see that the regions where the magnitude of the Poynting vector reaches maximum values are correlated to the concentration of vertical magnetic fluxes. 
The inclination of the magnetic field with respect to the vertical direction is depicted in Fig.~\ref{fig:uz+pf}(d). One can see that the magnitude of the Poynting vector is still considerable in some regions with higher inclination values, where one may expect upward energy flow according to Eq.\ \ref{eq:Pflux2}. Figure~\ref{fig:uz+pf}(e) displays the {magnitude of} the horizontal component of the velocity field, $|v_h|$. We also find some correspondence between regions with moderate to high Poynting flux and intense horizontal plasma motions, even in regions with weak vertical magnetic field. To evaluate the contribution of {the horizontal component, $\mathbf{S}_h=S_x \mathbf{i}+S_y \mathbf{j}$}, to the total Poynting flux, we write $\mathbf{S}^2 = \mathbf{S}_h^2+\mathbf{S}_z^2 $ 
and calculate the ratio ${S}_h^2/{S}^2$, as shown in Fig. \ref{fig:uz+pf}(f).
Our results show that, except for a minor fraction of the surface, the Poynting flux is flowing mainly in the horizontal direction. {By comparing Figures \ref{fig:uz+pf}(c) and \ref{fig:uz+pf}(f), we can see that the maximum electromagnetic energy flux in the photosphere flows parallel to the surface, not in the upward direction.}

\begin{figure*}[htp]
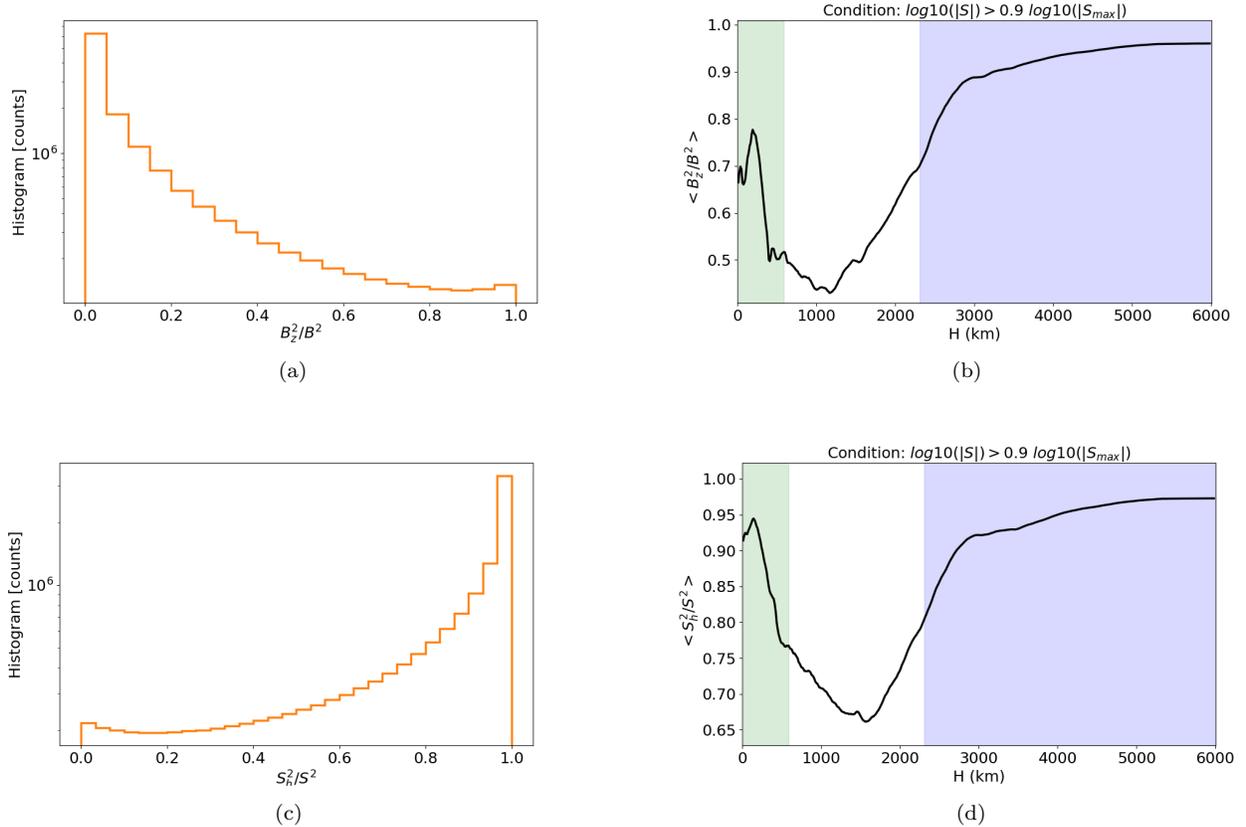

    \centering
    \gridline{
          \fig{Bzratio_dist3.png}{0.45\textwidth}{(a)}
          \fig{Bzratio_av.png}{0.45\textwidth}{(b)}
          }
\gridline{\fig{Shratio_dist2.png}{0.45\textwidth}{(c)}
          \fig{Shratio_av.png}{0.45\textwidth}{(d)}}

    \caption{Left panels: distribution of $B_z^2 /B^2$  (panel (a)), and the distribution of the ratio $S_h^2/S^2$ (panel (c)) from the solar surface up to a height of ~390 km. Right
panels: average value of $B_z^2 /B^2$ (panel (b)) and $S_h^2/S^2$ (panel (d)) for regions where $log_{10}(|S|)> 0.9 log_{10}(S_{max}) $ as a function of height. The colors indicate the average
height range of the photosphere (green), chromosphere (white), and transition region and corona (purple). }
    \label{fig:angles}
\end{figure*}

The distribution of the orientation of magnetic field at the photospheric level,{ i.e., up to 390 km from the solar surface}, is indicated in the histogram in Fig.~\ref{fig:angles}{(a)}, where we computed $B_z^2/B^2$. This histogram indicates that the simulated photospheric magnetic field has mostly high inclination. The vertical component of the magnetic field, $B_z$, is the principal component of the magnetic field in photospheric regions presenting concentrated flux of electromagnetic energy, as indicated in Fig.~\ref{fig:angles}{(b)}, where we plot the average value of $B_z^2/B^2$ in regions $log_{10}(|S|)> 0.9log_{10}(S_{max})$ as a function of height, {with  $S_{max}$ being the maximum value for Poynting flux at each height level}. Thereby, even though only a small portion of the magnetic field is aligned in the vertical direction, the Poynting flux is mainly originated by the vertical magnetic field in the lower photosphere. In contrast, closer to the chromosphere, the contribution of other magnetic components to the electromagnetic energy flux tends to increase. This indicates that most of the upflow of electromagnetic energy from the photosphere is originated slightly farther from the surface.  

 The dominant role of vertical magnetic field in contributing to the total electromagnetic energy flux indicates that the photospheric Poynting vector can be mostly retrieved from its horizontal component, ${\bf S}_h$. The horizontal nature of the Poynting vector is confirmed in Fig.\ref{fig:angles}{(c)}, where we plot the frequency distribution of the contribution of the horizontal Poynting vector to the total electromagnetic energy flux, $S_h^2/S^2$, for up to around 390 km above the solar surface. Fig.\ref{fig:angles}(c) shows that, close to the solar surface, only a tiny fraction of the electromagnetic energy generated by photospheric dynamics is flowing to the upper solar atmosphere or going back under the surface. The electromagnetic energy in the low photosphere will mostly flow within that height range. The $S_h^2/S^2$ frequency distribution compared to the $B_z^2/B^2$ histogram, panel (a), indicates that even magnetic fields with some inclination will likely not lead to upflows of electromagnetic energy. {In other words, in those cases the electromagnetic energy tends to flow perpendicular to the magnetic field along the plane defined by inclined magnetic fields}. In parts of the flow where we have considerable electromagnetic energy flux, $log_{10}(|S|)> 0.9log_{10}(S_{max})$, the  Poynting flux vector has a dominant horizontal component, as indicated by the average value of $S_h^2/S^2$ as a function of height displayed in Fig.~\ref{fig:angles}{(d)}.
 
 
\subsection{Defining an approximation to the Poynting vector}
The above findings indicate that in regions with high concentration of electromagnetic energy fluxes we can approximate the Poynting flux as
\begin{equation}
      \mathbf{S} \sim \mathbf{S}_{h}\sim  \frac{1}{4 \pi}  \mathbf{v}_{h}(\mathbf{B} \cdot \mathbf{B}) = \frac{1}{4 \pi}  \mathbf{v}_{h} (B_x^2+B_y^2+B_z^2).
     \label{eq:Pflux3}
\end{equation}
The approximation presented in Eq. ~\ref{eq:Pflux3} is still challenging to current observational data, as it depends on the horizontal component of the magnetic field. Despite the quiet-Sun area being also characterized by the transverse component of the magnetic field \citep{Jafarzadeh_2014,BellotRubio_2019}, these measurements are still limited owing to the limited spatial resolution and polarimetric accuracy of the present instrumentation. Figure~\ref{fig:angles}(a) shows that the contribution from the horizontal component of the magnetic field can be disregarded for some regions, especially for regions where $log_{10}(|S|)> 0.9 log_{10}(S_{max})$ as depicted in Fig.~\ref{fig:angles}(b). Therefore, we can further approximate the Eq.~\ref{eq:Pflux3} by
\begin{equation}
     \mathbf{S}^{obs} \sim   \frac{1}{4 \pi}  \mathbf{v}_{h} B_z^2,
     \label{eq:Pflux4}
\end{equation}
where $\mathbf{S}^{obs}$ is the observable Poynting flux, i.e., an estimation for Poynting flux based solely on quantities that can be easily retrieved from solar observational data. The above expression is rather similar to the approximation made before by {\citet{Fisher_1998} and \citet{Tan_2007}} when estimating the Poynting flux in active regions.
However, those studies actually applied the approximation in Eq.\ \ref{eq:Pflux4} to estimate the the electromagnetic energy flowing upward. 

\begin{figure*}[htp]
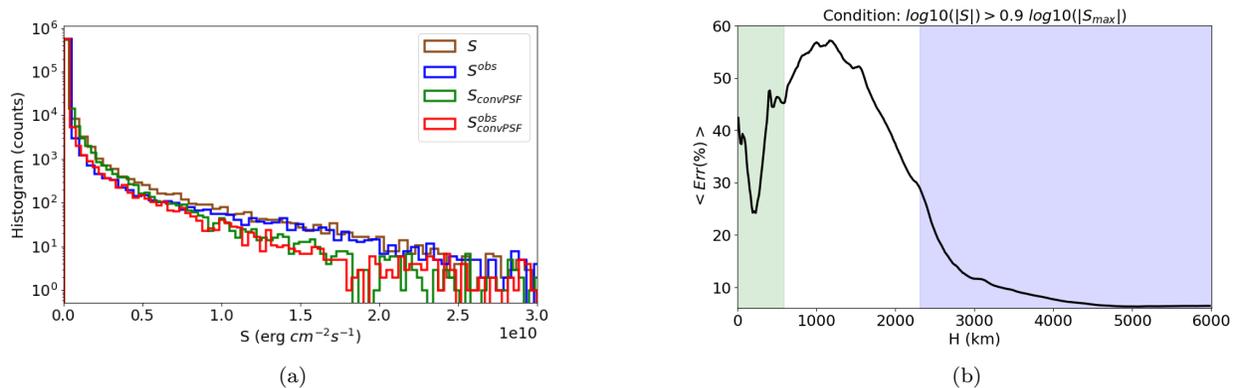

    \centering
    \gridline{
          \fig{P_simulazione_suzana.png}{0.45\textwidth}{(a)}
          \fig{err_av.png}{0.45\textwidth}{(b)}
          }
    \caption{(a)Distribution of the Poynting flux derived from the simulation as described by Eq.~\ref{eq:Pflux2} (blue histogram) and with the approximation (brown distribution) using the Eq.~\ref{eq:Pflux4}. The red/green distributions have been obtained by convolving the original simulation with the IMaX PSF.(b) Average value of the relative error, $Err$, for regions where $log_{10}(|S|)>0.9 log_{10}(S_{max})$ as a function of height.}
    \label{fig:erros}
\end{figure*}

To evaluate the validity of Eq.~\ref{eq:Pflux4}, we applied our approximation in conjunction with our simulation data and compared the magnitude of $\mathbf{S}^{obs}$ to the total magnitude of the Poynting flux obtained using Eq.~\ref{eq:Pflux}.
Figure~\ref{fig:erros}(a) displays the distribution of the Poynting flux derived from the simulation, as described by Eq.~\ref{eq:Pflux2} (brown histogram) and with the approximation  using Eq.~\ref{eq:Pflux4} (blue distribution). We find that the approximation described by Eq.~\ref{eq:Pflux4} is able to accurately capture the distribution of Poynting flux in the photosphere. For the regions where $log_{10}(|S|)>0.9 log_{10}(S_{max})$, the relative error between $S_{obs}$ and $S$ is, on average, lower than {$23.4\%$} in the photosphere as displayed in Figure~\ref{fig:erros}(b). Finally, for a better comparison between the simulation and the observations, we convolved the simulation with a PSF with a width comparable that of the IMaX instrument. These results are displayed in Fig.~\ref{fig:erros}(a) as green and red histograms, respectively. Our results show a great agreement for the approximated and total Poynting flux.

\subsection{Comparison with observational data}

\begin{figure*}[ht!]
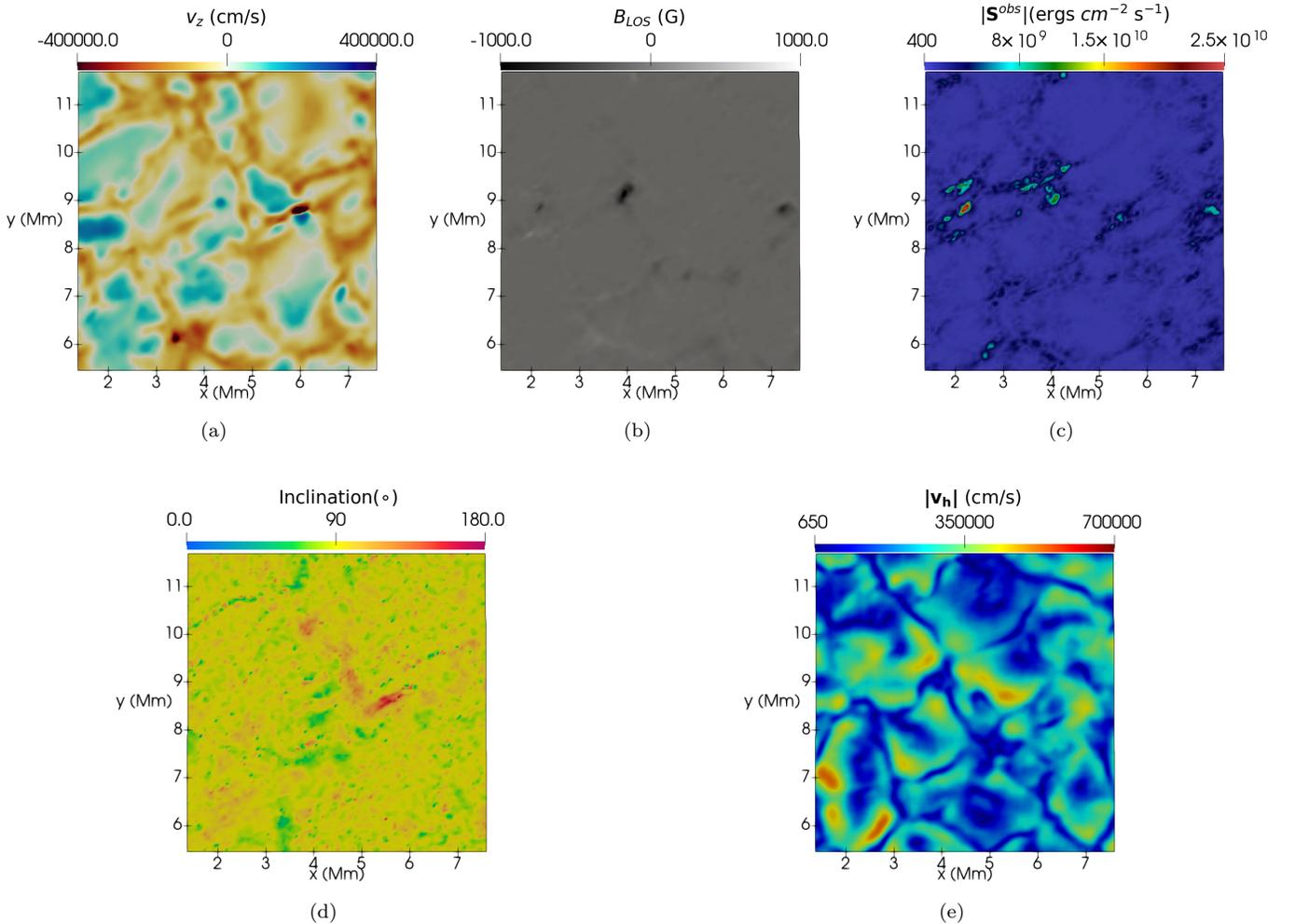

    \centering
\gridline{\fig{obs_vz.png}{0.33\textwidth}{(a)}
          \fig{obs_blong.png}{0.33\textwidth}{(b)}
\fig{obs_S.png}{0.33\textwidth}{(c)}}
\gridline{
          \fig{obs_inclination.png}{0.33\textwidth}{(d)}
          \fig{obs_vh.png}{0.33\textwidth}{(e)}          }
    \caption{Close view of a $\sim$6.2$\times$6.2 Mm region from the IMaX observation indicated by the white rectangle in Fig.~\ref{fig:IMaX}. (a) the vertical component of the velocity field; (b) the vertical component of the magnetic field, (c) the magnitude of the proxy of the Poynting flux; (d) the magnetic field inclination; (e) the horizontal plasma velocity. }
    \label{fig:deepvel}
\end{figure*}

Figure~\ref{fig:deepvel} displays a magnification of the {\sc Sunrise}/IMaX area inside the white cuboid shown in Fig.~\ref{fig:IMaX}. We have applied the Deepvel method to retrieve the horizontal plasma motion from observations and thereby compute the Poynting flux at each spatial position according to Eq.\ \ref{eq:Pflux4}. We display maps of the vertical components of the velocity and magnetic field (panels (a) and (b)), Poynting flux vector (panel (c)), magnetic field inclination (panel (d)), and horizontal component of the velocity field (panel (e)). 
These maps show that both the simulation and observation have a similar spatial distribution for the displayed variables, except for the inclination map. 
This discrepancy is caused by the limitations of inversion procedures.  This map is noisier than the simulation one with less spatial structuring owing to the noisy $Q$ and $U$ Stokes profiles in the quiet-Sun area. It is worth noting that the map in Fig. \ref{fig:deepvel}(c) shows that the observational photospheric Poynting flux spatial distribution is very similar to what was found in the simulation data. The Poynting flux tends to be intense and localized in small regions along the intergranular lanes, cospatial with line-of-sight magnetic field concentrations. The Deepvel horizontal velocity map displays a pattern resembling the granulation with values lower than those of the simulation data, even though the values are still greater than 3 km s$^{-1}$. 

\begin{figure*}[htp]
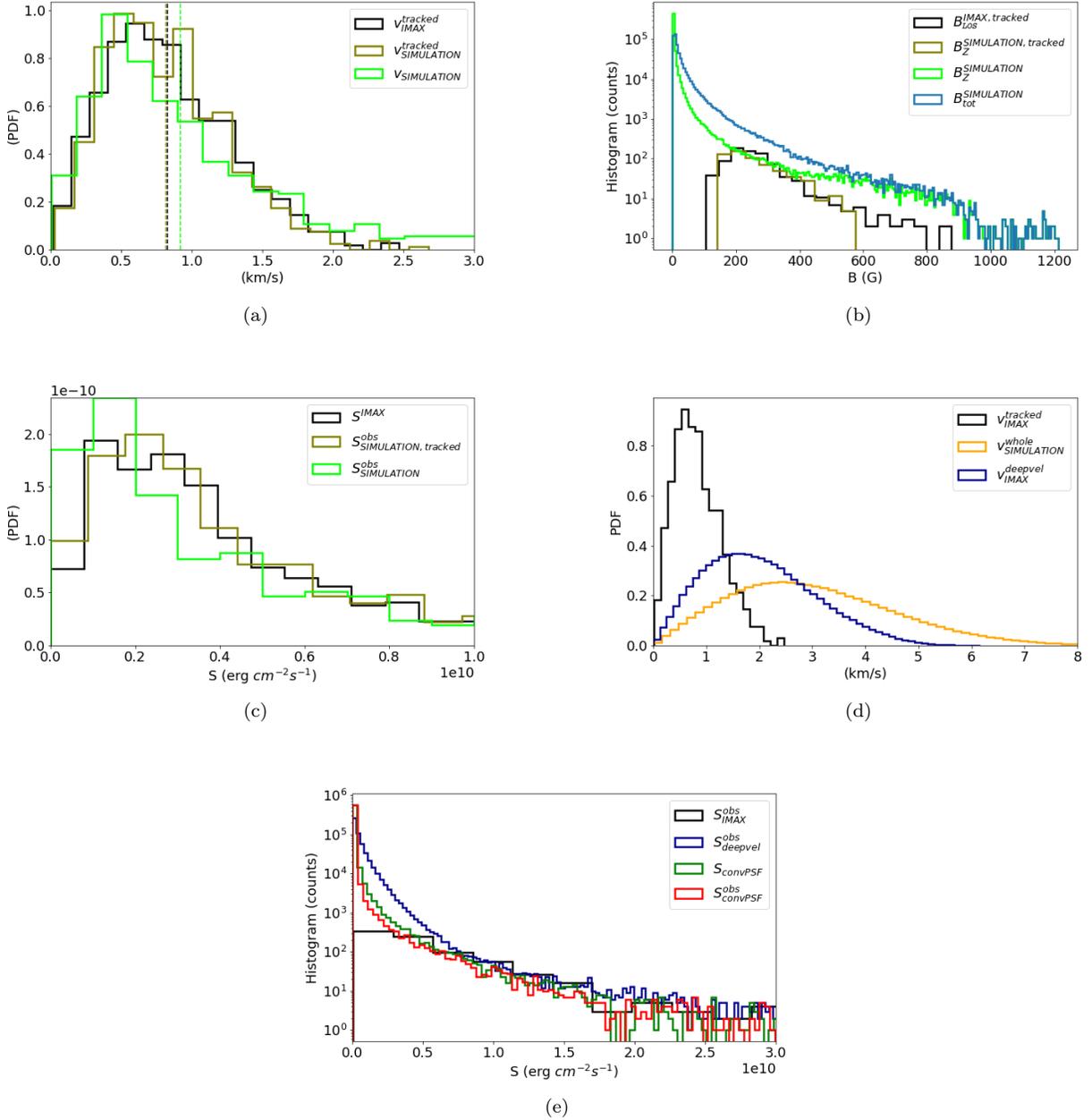

\centering
\gridline{
          \fig{velIMaXBifrostconvolvedtracking.png}{0.45\textwidth}{(a)}
          \fig{B_all.png}{0.45\textwidth}{(b)}
          }
\gridline{
          \fig{P_b_IMaX.png}{0.45\textwidth}{(c)}
          \fig{vel_all.png}{0.45\textwidth}{(d)}
          }
\gridline{
          \fig{P_simulazione_suzana_convolved+IMaX+deepvel.png}{0.45\textwidth}{(e)}
          }
    \caption{(a)Probability density function of the horizontal velocity field, $v_{h}$, associated with the magnetic elements tracked in the IMaX data (black histogram) and the simulation domain (olive histogram). The green distribution represents $v_{h}$ derived from the simulation at 115 km at the same magnetic elements selected. The three dashed lines indicate the average values of  $v_{h}$. (b) Distribution of the vertical magnetic field at the magnetic elements in the IMaX (black histogram) and the simulation (olive histogram) data sets. {For comparison, considering all pixels of the simulation domain at 115 km (i.e. photospheric heights), we also show $B_{z}$ and the total magnetic field (green and light blue histograms). } (c) Probability density function of the horizontal Poynting flux calculated by means of Eq.~\ref{eq:Pflux4}. (d) Overview of the horizontal velocity of the IMaX data (derived with two methods; black and blue histograms) and for the simulation (orange histogram). (e)  Distribution of the total Poynting flux using the Eq.~\ref{eq:Pflux3} (S$_{convPSF}$, green histogram), and the Poynting flux derived using Eq.~\ref{eq:Pflux4} (S$^{obs}_{convPSF}$, red curve). Both quantities are convolved with the IMaX PSF. The black histogram represents the Poynting flux as derived from the IMaX observations, and the blue curve shows the Poynting flux using the value of $v_{h}$ retrieved with the Deepvel method for the IMaX data.}
    \label{fig:vel_p_compare}
\end{figure*}

Through the feature tracking we obtained the horizontal velocity and the horizontal Poynting flux associated to each magnetic elements in both  the IMaX and Bifrost data. In particular, Fig.~\ref{fig:vel_p_compare} (a) displays the resulting horizontal velocity, $v_{h}$, showing that comparable results were found for  the $v_{h}$ retrieved using both simulation and {\sc Sunrise}/IMaX data. We also report for completeness the horizontal velocity of the simulation associated with each magnetic element at 115 km (see green histogram in Fig.~\ref{fig:vel_p_compare} (a)). For these elements we found a similar horizontal velocity on average between 0.5 and 1 km s$^{-1}$, in agreement with those reported in earlier studies \citep{Stangalini2013}. 

As the Poynting vector also depends on the magnetic field vector, in Fig.~\ref{fig:vel_p_compare}(b) we display the distribution of the magnetic field. Olive and black histograms represent the $z$-component of the magnetic field associated with each magnetic element for the simulation and the {\sc Sunrise}/IMaX data, respectively. The green and blue histograms indicate the $B_{z}$ component and the total magnetic field of the simulation at 115 km within the entire available FOV, respectively. As can be noted, when we applied the tracking features method, we do not consider two regions: those of the elements with small (less than 150 G) and high (greater than 800 G) values of the magnetic field. The lower limit is due to the smallest scale that our instrumentation can detect. On the other hand, the upper limit is the consequence of the diverse thresholds (spatial and temporal) that we set out. Another important aspect to be noted concerns the small difference between the simulation vertical and total component of the magnetic field. {Indeed, the intense magnetic field at photospheric heights is mostly vertical (see the green and blue curves)}. This further supports the application of our approximation discussed in Sect.~3.2. 

The plots displayed in Fig.~\ref{fig:vel_p_compare}(c) show the distributions of the Poynting vector when we consider the tracked magnetic elements. The distributions present a good agreement, which is expected since we found a comparable horizontal velocity and vertical magnetic field distributions for observations and simulations. Again, for completeness, in panel (c) we plot the Poynting vector derived using the horizontal velocity at 115 km of the simulation associated with each magnetic feature found (green histogram). In Fig.~\ref{fig:vel_p_compare} {(d)} we compare the horizontal velocity retrieved with the two methods (magnetic tracking and Deepvel) of the {\sc Sunrise}/IMaX data and the results of simulation. The horizontal velocity associated with the magnetic elements  peaks at around 0.8 km s$^{-1}$, while this result is shifted to higher values when we consider the whole FOV. 

Figure~\ref{fig:vel_p_compare} {(e)} displays the Poynting vector at each pixel of the whole simulation domain convolved with an opportune PSF with width similar to the IMaX instrument (red histogram) compared to that at the {\sc Sunrise}/IMaX magnetic features tracked (black histogram). This plot reveals a good agreement between the two data considered. Furthermore, for comparison, we also overplot the Poynting vector derived using  Eq.~\ref{eq:Pflux2} and the Poynting vector calculated using the Deepvel horizontal velocity at each available spatial position (green and blue histograms, respectively). Except for the low-energy part, all the computations for Poynting vector present a similar behavior. The Poynting vector of the {\sc Sunrise}/IMaX data ( black histogram) appears flatted at small energy, probably due to the lower limit of detected signal below our capabilities. In this low-energy range the Poynting vector calculated using the Deepvel horizontal velocity has greater values owing to the overestimated horizontal velocity field, intrinsic of this method \citep{Asensio2017}. Indeed, the neural network was trained with IMaX data with 30 s cadence, instead of the 33 s of the original IMaX cadence. \citet{Asensio2017} report DeepVel velocities 1.15 times larger in magnitude than the Local  Correlation  Tracking methodology.

\section{Discussions and conclusions}
\begin{figure*}
    \centering
    \includegraphics[width=0.48\textwidth]{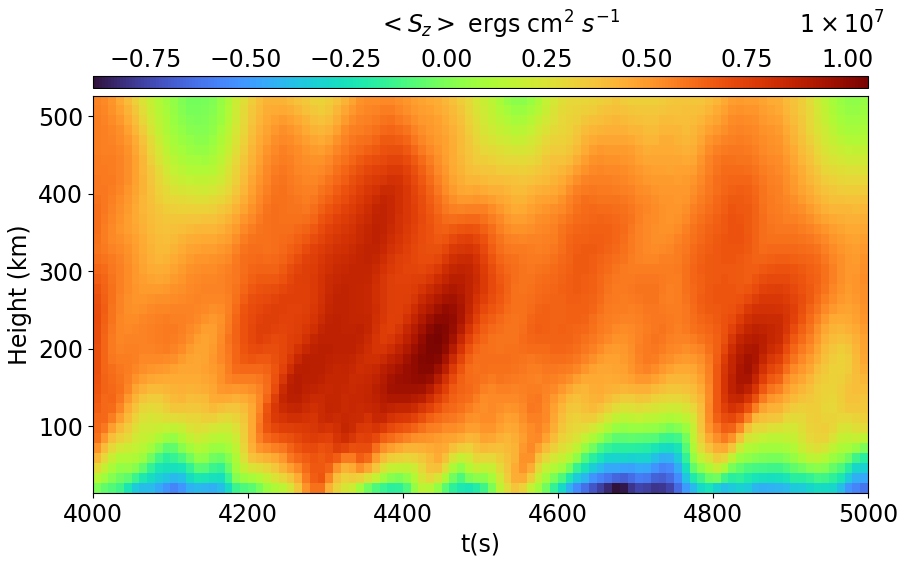}
        \includegraphics[width=0.33\textwidth]{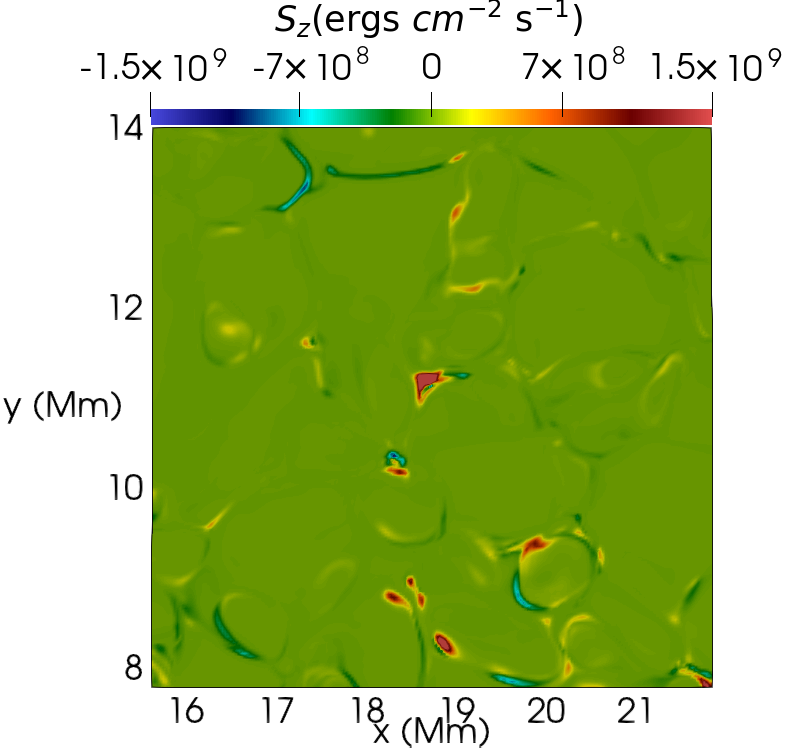}
    \caption{{Left panel: the average value of $S_z$ for each height level above the surface, up to 500 km, and for each time frame within a time interval of 1000 s (16.66 minutes). The vertical and horizontal axes indicate the height and time, respectively. Right panel: close view of a 6.25 $\times$ 6.25 Mm$^{2}$ region from Bifrost simulation indicated by the white rectangle in Fig. \ref{fig:simulation_domain}. The horizontal plane is placed around 100 km above the surface, and it is colored by the $z$-component of the Poynting flux, $S_z$}}
    \label{fig:verticalS}
\end{figure*}

The electromagnetic energy flow in the lower atmosphere is an essential aspect to describe the dynamics and energy balance of the solar atmosphere. Estimating the energy flux density accurately from observations and determining the direction of that energy flux is crucial for describing accurately the process of plasma heating. Based on realistic simulations from Bifrost, we find that the electromagnetic energy flux occurs practically along the horizontal plane for most of the photospheric region. 
The dominance of the horizontal component of the energy flux is observed in both the photosphere and other atmospheric layers, being especially higher in regions where we find strong electromagnetic energy concentration and where the magnetic field is mostly vertical. {However, our findings do not preclude the existence of a vertical component capable of transporting enough energy to justify chromospheric heating. The left panel of Fig. \ref{fig:verticalS} displays the average of the vertical component of the Poynting flux, $<S_z>$, for each height level above the surface and for each time frame within a time interval of 1000 s. For heights of 100 km above the solar surface, the net vertical Poynting flux provides a positive energy input for the upper solar atmosphere. As the radiative losses in a chromospheric quiet-Sun region are estimated to be around $4\times 10^{6}$ ergs cm$^{-2}$ s$^{-1}$ \citep[e.g]{Withbroe_1977,Vernazza_1981}, we find that the net $S_z$ is enough to compensate chromospheric losses for 100$\leq H \leq 400$ km. The right panel of Fig.\ref{fig:verticalS} depicts the distribution of the vertical Poynting flux for the Bifrost simulation at the same height and time as Figure \ref{fig:uz+pf}, H$\sim$ 115 km km and t=76' 46''. Comparing Fig. \ref{fig:uz+pf}(c) and the right panel of Fig. \ref{fig:verticalS}, it is clear that the higher values of vertical Poynting flux are found in the same regions with a local maximum of the total electromagnetic energy flux. Therefore, the estimate of our proxy for Poynting flux can also provide a good estimate of the regions with higher contributions for energy transport in the vertical direction.}

As most of the electromagnetic energy is flowing horizontally, energy can be concentrated by vortices as indicated by \cite{Shelyag_2012}. As the swirling motion perturbs the magnetic lines, it can create a net upward Poynting flux \citep{Shelyag_2012,Yadav_2020}. {Our results indicate that a magnetic field with a high inclination does not necessarily lead to a dominant vertical component of the Poynting flux. The vertical component of the Poynting vector tends to contribute more to the total energy flux in the upper part of the photosphere and the chromosphere. However, the upper photospheric regions do not necessarily present higher vertical Poynting flux when compared to regions around 100 km above the solar surface. Therefore, the dominance of the horizontal electromagnetic energy flux does not imply a lower vertical energy flow.}
In a recent study \cite{Battaglia_2021} indicated that the upper photosphere and chromospheric swirls can provide a positive energy input. 

Based on the horizontal nature of the Poynting vector, we established an approximation for the Poynting flux that requires only the horizontal plasma motion and the vertical component of the magnetic field, quantities retrieved from observations. In other words, our approximation describes the Poynting vector as the magnetic energy from vertical magnetic flux flowing in the direction of the horizontal plasma flow. Our proxy for the Poynting vector can predict the actual Poynting vector magnitude with an underestimation {factor less than $23.4\%$ on average for regions with intense electromagnetic energy flux in the photosphere (100$\leq H \leq 300$ km). In other words, the proxy provides a lower limit for the observed Poynting flux.
The height extension where our proxy works with a lower underestimation factor is within what is possible to retrieve from current available data observations. A greater height range analysis would require multiline polarimetric data, which are not available for the quiet Sun at the moment. This aspect could be theoretically analyzed and might constitute the subject for future investigations.}

 We found a similar Poynting flux proxy distribution for both simulated and observational data, with maximum values around 10$^{10}$ ergs cm$^{-2}$ s$^{-1}$. The {\sc Sunrise}/IMaX data  reach considerably larger values for the Poynting flux when compared to previous estimates found for vertical electromagnetic energy flux based on high-resolution observations of plage magnetic fields  \citep{Yeates_2014,Welsch_2015}. This is a strong indication that the photospheric electromagnetic energy flux tends to be mainly horizontal in quiet-Sun regions, as our proxy is based on the horizontal component of the Poynting flux. Although our results for the upper atmosphere strongly depend on the magnetic field and plasma flow conditions, our results describe the Poynting vector for magnetic network patches and solar plage regions, where the magnetic field is primarily vertical. The fact that the Poynting flux has a considerably small vertical component in the photospheric region indicates that, even for other magnetic field configurations, it is likely that we are greatly underestimating the total electromagnetic energy flow by only considering the vertical component. Failing to correctly estimate the full Poynting flux in other atmospheric regions  may lead to the wrong description of energy transport.  Our results suggest that future investigations should concern the full Poynting vector to evaluate its contribution to local heating properly.

\acknowledgments
VF, GV, IB and SSAS are grateful to The Royal Society, International Exchanges Scheme, collaboration with Brazil (IES$\backslash$R1$\backslash$191114) and Spain (IES$\backslash$R2$\backslash$212183). VF, GV and SSAS are grateful to Science and Technology Facilities Council (STFC) grant ST/V000977/1 and to The Royal Society, International Exchanges Scheme, collaboration with Chile (IE170301). VF would like to thank the International Space Science Institute (ISSI) in Bern, Switzerland, for the hospitality provided to the members of the team on ‘The Nature and Physics of Vortex Flows in Solar Plasmas’. This research has also received financial support from the European Union’s Horizon 2020 research and innovation program under grant agreement No. 824135 (SOLARNET). SJ acknowledges support from the European Research Council under the European Union Horizon 2020 research and innovation program (grant agreement No. 682462) and from the Research Council of Norway through its Centres of Excellence scheme (project No. 262622).
The German contribution to {\sc Sunrise} is funded by the Bundesministerium f\"{u}r Wirtschaft und Technologie through the Deutsches Zentrum f\"{u}r Luft- und Raumfahrt e.V. (DLR), grant No. 50 OU 0401, and by the Innovations fond of the President of the Max Planck Society (MPG). The Spanish contribution has been funded by the Spanish MICINN under projects ESP2006-13030-C06 and AYA2009-14105-C06 (including European FEDER funds). The HAO contribution was partly funded through NASA grant NNX08AH38G.
The Bifrost simulations have been run on clusters from the Notur project, and the Pleiades cluster through the computing project s1061, s2053 and s8305 from the High End Computing (HEC) division of NASA. This study was financed in part by the Coordenação de Aperfeiçoamento de Pessoal de Nível Superior – Brasil (CAPES) – Finance Code 88882.316962/2019-01
MM aknowledges support from the European Union’s Horizon 2020 Research and Innovation 531 program under grant agreements No 824135 (SOLARNET) and No 729500 (PRE-EST) and from the Italian MIUR-PRIN grant 2017 "Circumterrestrial environment: Impact of Sun-Earth Interaction" and by the Istituto NAzionale di Astrofisica (INAF).

\bibliography{bibliografia}{}
\bibliographystyle{aasjournal}

\end{document}